\documentclass[12pt]{article}
\def\al{\alpha}
\def\be{\beta}
\def\ga{\gamma}
\def\de{\delta}
\def\ep{\epsilon}

\def\ze{\zeta}
\def\et{\eta}

\def\ka{\kappa}
\def\la{\lambda}

\def\rh{\rho}

\def\si{\sigma}

\def\ph{\phi}

\def\ch{\chi}
\def\ps{\psi}

\def\De{\Delta}

\def\mn{{\mu\nu}}
\def\cl{{\mathcal L}}
\def\fr#1#2{{{#1} \over {#2}}}
\def\prt{\partial}

\def\half{{\textstyle{1\over 2}}}
\def\frac#1#2{{\textstyle{{#1}\over {#2}}}}

\def\lsim{\mathrel{\rlap{\lower4pt\hbox{\hskip1pt$\sim$}}
    \raise1pt\hbox{$<$}}}
\def\gsim{\mathrel{\rlap{\lower4pt\hbox{\hskip1pt$\sim$}}
    \raise1pt\hbox{$>$}}}
\def\sqr#1#2{{\vcenter{\vbox{\hrule height.#2pt
         \hbox{\vrule width.#2pt height#1pt \kern#1pt
         \vrule width.#2pt}
         \hrule height.#2pt}}}}

\def\etal {{\it et al.}}
\newcommand{\beq}{\begin{equation}}
\newcommand{\eeq}{\end{equation}}
\newcommand{\bea}{\begin{eqnarray}}
\newcommand{\eea}{\end{eqnarray}}
\newcommand{\rf}[1]{(\ref{#1})}

\def\uc#1{\uppercase{#1}}

\begin{document}

\title{
Astrophysical Tests of Lorentz Symmetry in Electrodynamics
\footnote{
invited talk presented at
\uc{F}ourth \uc{I}nternational \uc{W}orkshop on
\uc{N}ew \uc{W}orlds in \uc{A}stroparticle \uc{P}hysics,
\uc{U}niversity of the \uc{A}lgarve,
\uc{F}aro, \uc{P}ortugal, \uc{S}eptember, 2002.}
}

\author{MATTHEW MEWES\\
  Physics Department, Indiana University,\\
  Bloomington, IN 47405, U.S.A.\\
  E-mail: mmewes@indiana.edu} 

\date{}

\maketitle

\begin{abstract}
In this talk presented at
the Fourth International Workshop on
New Worlds in Astroparticle Physics,
I discuss recent constraints on Lorentz
violation in electrodynamics.
The observed absence of birefringence
of light that has propagated over cosmological
distances bounds some coefficients for
Lorentz violation to $2\times 10^{-32}$.
\end{abstract}

\section{Introduction}
The exact character of physics beyond the
standard model is an open question.
The standard model is commonly believed to
be the low-energy limit of Planck-scale physics
which unifies all known forces.
Due to the energy scales involved,
an experimental search for this new physics
would seem pointless given the current
technology.
However, some high-energy theories
may lead to violations in symmetries
which hold exactly in the standard model
\cite{kps,ncqed}.
In particular, spontaneous symmetry breaking
in the fundamental theory might result in apparent
violations in the Lorentz and CPT symmetries.
Furthermore, Lorentz and CPT violations
can be tested to extremely high precision
using today's technology
\cite{cpt01}.

A general Lorentz-violating
extension to the standard model
has been constructed
\cite{cpt01,ck}.
It consists of the minimal standard model
plus small Lorentz- and CPT-violating terms.
The standard-model extension has provided a
theoretical framework for many searches for
Lorentz and CPT violations.
To date, experiments involving
hadrons
\cite{hadronexpt,hadronth},
protons and neutrons
\cite{pn},
electrons
\cite{eexpt,eexpt2},
photons
\cite{cfj,km},
and muons
\cite{muons}
have been performed.

In practice, one often works with
a particular limiting theory
extracted from the standard-model extension.
For example, the photon sector of the
standard-model extension
yields a Lorentz-violating modified
electrodynamics.
The theory predicts several
unconventional features that lead to
sensitive tests of Lorentz symmetry.
For example, in the presence of certain
forms of Lorentz violation, light
propagating through the vacuum
will experience birefringence.
The absence of birefringence
in light emitted from distant sources
leads to tight bounds on some of the
coefficients for Lorentz violation
\cite{cfj,km}.

In this work, I review some of these bounds.
This research was done in collaboration
with Alan Kosteleck\'y.
A detailed discussion can be found
in the literature
\cite{km}.

\section{Lorentz-Violating Electrodynamics}
The modified electrodynamics
maintains the usual gauge invariance
and is covariant under observer
Lorentz transformations.
It includes both CPT-even and -odd terms.
The CPT-odd terms have been the
subject of numerous experimental and
theoretical investigations
\cite{ck,cfj,jk,klp}.
For example, some of these terms have
been bounded to extremely high precision
using polarization measurements
of distant radio galaxies
\cite{cfj}.
In contrast, until recently,
the CPT-even terms have received
little attention.
Here, I review a recent study of
these terms
\cite{km}.

The CPT-even lagrangian for
the modified electrodynamics is
\cite{ck}
\beq
\cl=-\frac14 F_{\mu\nu}F^{\mu\nu}
-\frac14 (k_F)_{\ka\la\mu\nu}
     F^{\ka\la}F^{\mu\nu}\ ,
\label{lag}
\eeq
where $F_\mn$ is the field strength,
$F_\mn \equiv \prt_\mu A_\nu -\prt_\nu A_\mu$.
The first term is the usual
Maxwell lagrangian.
The second is an unconventional
Lorentz-violating term.
The coefficient for
Lorentz violation,
$(k_F)_{\ka\la\mu\nu}$,
is real and comprised of
19 independent components.
The absence of observed
Lorentz violation implies
$(k_F)_{\ka\la\mu\nu}$ is small.
The equations of motion for
this lagrangian are 
$\prt_\al{F_\mu}^\al
+(k_F)_{\mu\al\be\ga}\prt^\al F^{\be\ga}=0$.
These constitute modified source-free
inhomogeneous Maxwell equations.
The homogeneous Maxwell equations
remain unchanged.

A particularly useful
decomposition of the 19 independent
components can be made
\cite{km}.
The lagrangian in terms of this
decomposition is 
\bea
\cl&=&\half[(1+\tilde\ka_{\rm tr})\vec E^2
-(1-\tilde\ka_{\rm tr})\vec B^2]
+\half \vec E\cdot(\tilde\ka_{e+}
+\tilde\ka_{e-})\cdot\vec E
\nonumber\\
&&
-\half\vec B\cdot(\tilde\ka_{e+}
-\tilde\ka_{e-})\cdot\vec B 
+\vec E\cdot(\tilde\ka_{o+}
+\tilde\ka_{o-})\cdot\vec B\ ,
\label{lag2}
\eea 
where  $\vec E$ and $\vec B$ are
the usual electric and magnetic fields.
The $3\times3$ matrices
$\tilde\ka_{e+}$, $\tilde\ka_{e-}$,
$\tilde\ka_{o+}$ and $\tilde\ka_{o-}$
are real and traceless.
The matrix $\tilde\ka_{o+}$ is antisymmetric,
while the remaining three are symmetric.
The real coefficient
$\tilde\ka_{\rm tr}$
corresponds to the only rotationally
invariant component of $(k_F)_{\mu\al\be\ga}$.

From the form of Eq.~\rf{lag2},
we see that the component
$\tilde\ka_{\rm tr}$
can be thought of as a shift in the
effective permittivity $\ep$ and
effective permeability $\mu$ by
$(\ep-1)=-(\mu^{-1}-1)
=\tilde\ka_{\rm tr}$.
The result of this shift
is a shift in the speed of light.
Normally, this may be viewed
as a distortion of the metric.
In fact, this result generalizes
to the nine independent coefficients in
$\tilde\ka_{\rm tr}$,
$\tilde\ka_{e-}$ and
$\tilde\ka_{o+}$.
To leading order,
these may be viewed as a distortion
of the spacetime metric of the form
$\et^\mn\rightarrow\et^\mn+k^\mn$,
where $k^\mn$ is small, real and
symmetric.

Small distortions of this type
are unphysical, since they can
be eliminated through
coordinate transformations and
field redefinitions.
However, each sector
of the full standard-model
extension contains similar
terms.
Eliminating these terms from
one sector will alter the
other sectors.
Therefore, the effects of such terms
can not be removed completely from
the theory.
As a consequence, in experiments where
the properties of light are
compared to the properties of
other sectors,
these terms are relevant.
However, in experiments where
only the properties of light
are relevant,
the nine coefficients in
$\tilde\ka_{\rm tr}$,
$\tilde\ka_{e-}$ and
$\tilde\ka_{o+}$
are not expected to appear.
The tests discussed here
rely on measurements of
birefringence.
This involves comparing
the properties of light
with different polarizations.
Therefore, these tests compare 
light with light and are only
sensitive to the ten independent
components of $\tilde\ka_{e+}$
and $\tilde\ka_{o-}$.

Constraints on birefringence
have been expressed
in terms of a ten-dimensional
vector $k^a$ containing the ten
independent components of
$\tilde\ka_{e+}$ and $\tilde\ka_{o-}$
\cite{km}.
The relationship between
$\tilde\ka_{e+}$, $\tilde\ka_{o-}$
and $k^a$ is given by
\bea
(\tilde\ka_{e+})^{jk} &=&
-\left(
\begin{array}{ccc}
-(k^3+k^4) & k^5 & k^6 \\
k^5 & k^3 & k^7 \\
k^6 & k^7 & k^4 
\end{array}
\right)\ , 
\nonumber \\
(\tilde\ka_{o-})^{jk} &=&
\left(
\begin{array}{ccc}
2k^2 & -k^9 & k^8 \\
-k^9 & -2k^1 & k^{10} \\
k^8 & k^{10} & 2(k^1-k^2) 
\end{array}
\right)\ .
\eea
Bounds on birefringence appear as
bounds on $|k^a|\equiv\sqrt{k^ak^a}$,
the magnitude of the vector $k^a$.

\section{Birefringence}
In order to understand the
effects of Lorentz violation
on the propagation of light,
we begin by considering
plane-wave solutions.
Adopting the ansatz
$F_\mn(x)=F_\mn e^{-ip_\al x^\al}$
and solving the modified Maxwell
equations yields the
dispersion relation
\beq
p^0_\pm=(1+\rh\pm\si)|\vec p|\ .
\label{dispersion}
\eeq
In a frame where 
the phase velocity is along
the $z$-axis, the electric field
takes the form
\beq
\vec E_\pm\propto(\sin\xi,\pm1-\cos\xi,0)
+O(k_F)\ .
\label{efield}
\eeq
To leading order,
the quantities $\rh$,
$\si\sin\xi$ and $\si\cos\xi$
are linear combinations of
$(k_F)_{\ka\la\mu\nu}$
and depend on $\hat v$,
the direction of propagation.

A prediction of these solutions
is the birefringence of light in
the vacuum.
Birefringence is commonly found in
conventional electrodynamics 
in the presence of anisotropic media.
In the present context,
the general vacuum solution is
a linear combination of the
$\vec E_+$ and $\vec E_-$.
For nonzero $\si$, these solutions
obey different dispersion relations.
As a result, they propagate
at slightly different velocities.
At leading order, the difference
in the velocities is given by
\beq
\De v \equiv v_+-v_- = 2\si\ .
\label{dev}
\eeq
For light propagating over
astrophysical distances, this
tiny difference may become apparent.

As can be seen from the
above solutions,
birefringence depends on
the linear combination
$\si\sin\xi$ and $\si\cos\xi$.
As expected, these only contain the ten
independent coefficients which appear
in $\tilde\ka_{e+}$ and $\tilde\ka_{o-}$.
Expressions for 
$\si\sin\xi$ and $\si\cos\xi$
in terms of these ten
independent coefficients and
the direction of propagation
can be found in the literature
\cite{km}.

Next I discuss two observable
effects of birefringence.
The first effect is the spread
of unpolarized pulses of light.
The second is the change in
the polarization angles of
polarized light.

\subsection{Pulse-Dispersion Constraints}
The narrow pulses of radiation
from distant sources such as pulsars
and gamma-ray bursts are well
suited for searches for birefringence.
In most cases, the pulses are 
relatively unpolarized.
Therefore, the components $\vec E_\pm$
associated with each mode will be comparable.
The difference in velocity 
will induce a difference in the
observed arrival time of the two modes
given by $\De t \simeq \De v L$,
where $L$ is the distance to the source.

Sources which produce radiation with
rapidly changing time structure may
be used to search for this difference
in arrival time.
For example, the sources mentioned above
produce pulses of radiation.
The pulse can be regarded as the
superposition of two independent
pulses associated with each mode.
As they propagate, the
difference in velocity will cause
the two pulses to separate.
A signal for Lorentz violation would
then be a measurement of two sequential
pulses of similar time structure.
The two pulses would be linearly
polarized at mutually orthogonal
polarization angles.

The above signal for birefringence
has not yet been observed.
However, existing pulse-width measurements 
place constraints on Lorentz violation.
To see this, suppose a source produces a
pulse with a characteristic time width $w_s$.
As the pulse propagates, the two
modes spread apart and the width of the
pulse will increase.
The observed width can be estimated
as $w_o \simeq w_s+\De t$.
Therefore, observations of $w_o$
place conservative bounds on
$\De t \simeq \De v L \simeq 2\si L$.
The resulting bound on $\si$ constrains
the ten-dimensional parameter space of
$\tilde\ka_{e+}$ and $\tilde\ka_{o-}$.
Since a single source constrains only
one degree of freedom, at least ten
sources located at different positions
on the sky are required to fully constrain
the ten coefficients.

Using published pulse-width measurements
for a small sample of fifteen pulsars
and gamma-ray bursts, we found
bounds on $\si$ for fifteen different
propagation directions $\hat v$.
Combining these bounds constrained
the ten-dimensional parameter space.
At the 90\% confidence level, we
obtained a bound of $|k^a| < 3 \times 10^{-16}$
on the coefficients for Lorentz violation
\cite{km}.

\subsection{Polarimetry Constraints}
The difference in the velocities of
the two modes results in changes
in the polarization of polarized light.
Decomposing a general electric field
into its birefringent components,
we write
$\vec E(x) = (\vec E_+ e^{-ip^0_+t}+
\vec E_- e^{-ip^0_-t})e^{i\vec p \cdot\vec x}$.
Each component propagates
with a different phase velocity.
Consequently, the relative phase
between modes changes as the light
propagates.
The shift in relative phase
is given by
\beq
\De\ph
= (p^0_+-p^0_-)t 
\simeq 4\pi\si L/\la\ ,
\eeq
where $L$ is the distance to the
source and $\la$ is the wavelength
of the light.
This phase change results in
a change in the polarization.

The $L/\la$ dependence suggests
the effect is larger for more
distant sources and shorter
wavelengths.
Recent spectropolarimetry of distant
galaxies at wavelengths ranging from
infrared to ultraviolet
has made it possible to achieve values
of $L/\la$ greater than $10^{31}$.
Given that measured polarization parameters
are typically of order 1,
we find an experimental
sensitivity of $10^{-31}$ or better
to components of $(k_F)_{\ka\la\mu\nu}$.

In general, plane waves
are elliptically polarized.
The polarization ellipse can
be parameterized with angles 
$\psi$, which characterizes the
orientation of the ellipse, and
$\chi=\pm\arctan
\frac{\rm minor\ axis}{\rm major\ axis}$,
which describes the shape of the
ellipse and helicity of the wave.
The phase change, $\De\ph$, results
in a change in both $\psi$ and $\chi$.
However, measurements of $\chi$ are not
commonly found in the literature.
Focusing our attention on $\psi$,
we seek an expression for
$\de\psi=\psi-\psi_0$,
the difference between $\psi$ at
two wavelengths, $\la$ and $\la_0$.
We find
\cite{km}
\beq
\de\psi
=\half\tan^{-1}{\fr
{\sin\tilde\xi\cos\ze_0
+\cos\tilde\xi\sin\ze_0\cos(\de\ph-\ph_0)}
{\cos\tilde\xi\cos\ze_0
-\sin\tilde\xi\sin\ze_0\cos(\de\ph-\ph_0)}},
\label{dpsi}
\eeq
where 
$\de\ph=4\pi\si(L/\la-L/\la_0)$,
$\tilde\xi=\xi-2\psi_0$ and
$\ph_0 \equiv \tan^{-1}
(\tan2\ch_0/\sin\tilde\xi)$,
$\ze_0\equiv \cos^{-1}
(\cos2\ch_0\cos\tilde\xi)$.
The polarization at $\la_0$ is given
by the polarization angles
$\ps_0$ and $\ch_0$.

The idea is to fit
existing spectropolarimetric
data to Eq.~\rf{dpsi}.
Under the reasonable assumption
that the polarization of the
light when emitted is relatively
constant over the relevant wavelengths,
any measured wavelength dependence
in the polarization is due to Lorentz
violation.
Using a sample of sixteen distant galaxies
with published polarimetric data
with wavelengths ranging from 400 to 2200 nm,
we were able to place a constraint
on the ten coefficients in 
$\tilde\ka_{e+}$ and $\tilde\ka_{o-}$.
As in the pulse-dispersion case,
we first determined a bound on $\si$
for each source.
Combining these bounds, we obtain
a constraint on the ten-dimensional
parameter space of $k^a$.
At the 90\% confidence level,
this gave a bound of
$|k^a| < 2 \times 10^{-32}$
on the ten coefficients for
Lorentz violation responsible
for birefringence
\cite{km}.

\section{Summary}
I have reviewed a recent study of
Lorentz violation in electrodynamics
\cite{km}.
I have described how pulse-width measurements
lead to a constraint of
$3 \times 10^{-16}$ on Lorentz violation.
I have also discussed how measurements
of polarization angles of light emitted
from galaxies at cosmological distances
lead to a constraint of
$2 \times 10^{-32}$
on ten coefficients for
Lorentz violation.

\section*{Acknowledgments}
I thank Alan Kosteleck\'y for
collaboration on this work.

\end{document}